\documentclass[acus]{JAC2000}

\usepackage{graphicx}

\newcommand{\Cleoc}{CLEO-c}

\newcommand{\Jpsi}{J/\psi}
\newcommand{\ie}{{\it i.e.}}
\newcommand{\BES}{BES Collaboration}

\newcommand{\Begitem}{\begin{enumerate}}
\newcommand{\Enditem}{\end{enumerate}}

\newcommand{\fbinvv}	{\mbox{${\rm fb}^{-1}$}}
\newcommand{\lumi}[1]   {\mbox{$\times 10^{#1}~{\rm cm}^{-2}{\rm s}^{-1}$}}
\newcommand{\dbar}	{\mbox{${\overline{D}}$}}
\newcommand{\dsbar}	{\mbox{${\overline{D_s}}$}}
\newcommand{\ds}	{\mbox{${D_s}$}}
\newcommand{\dd}	{\mbox{${D}$}}

\newcommand{\avcd}	{\mbox{$|V_{cd}|$}}
\newcommand{\avtd}	{\mbox{$|V_{td}|$}}

\newcommand{\avcs}	{\mbox{$|V_{cs}|$}}
\newcommand{\avts}	{\mbox{$|V_{ts}|$}}

\newcommand{\avub}	{\mbox{$|V_{ub}|$}}
\newcommand{\avcb}	{\mbox{$|V_{cb}|$}}

\newcommand{\fd}	{\mbox{$f_{D}$}}
\newcommand{\fds}	{\mbox{$f_{D_s}$}}
\newcommand{\fb}	{\mbox{$f_{B}$}}
\newcommand{\fbs}	{\mbox{$f_{B_s}$}}

\newcommand{\qsq}	{\mbox{$q^2$}}

\begin{document}
\title{\flushright{T06}\\[15pt] \centering THE PROPOSED CLEO-C PROGRAM AND R MEASUREMENT PROSPECTS\thanks{Adapted from "CLEO-c and CESR-c:
 A New Frontier of Weak and Strong Interactions", the CLEO-c project description,
by the CLEO Collaboration}}

\author{presented by L. Gibbons, Cornell University, Ithaca, NY, 14850, USA}

\maketitle

\begin{abstract}
The proposed experimental program (\Cleoc) for a charm factory 
based on a modification of the Cornell Electron Storage Ring
is summarized.   The prospects for $R$ measurements over
the range 3 GeV $\le \sqrt{s} \le$ 7 GeV are examined in detail.
\end{abstract}

\section{CLEO-c Program Overview}

The CLEO collaboration is proposing a focused three-year program of 
charm and QCD physics with
the CLEO detector operating in the range $\sqrt{s}=3-5$ GeV.   The
\Cleoc\ physics program includes a set of measurements that will
substantially advance our understanding of important Standard Model
processes and set the stage for understanding the larger theory in which
we imagine the Standard Model to be embedded.

Much of this program revolves around the strong interactions and the
pressing need to develop sufficiently powerful tools to deal with an
intrinsically non-perturbative theory.  At the present time, and for the
last twenty years, progress in weak interaction physics and the study of
heavy flavor physics has been achieved primarily by seeking those few
probes of weak-scale physics that successfully evade or minimize the
role of strong interaction physics.  The pre\"{e}minence of the mode
$B\to J/\psi K_S$ in measuring $\sin2\beta$ stems almost entirely from
the absence of  complications due to the strong interactions.  
Similarly, the discovery of a previously
unrecognized symmetry in QCD, which led to Heavy Quark Effective Theory
(HQET), created an opportunity in heavy-to-heavy quark decays where
strong interaction effects are minimized.  HQET's identification of the
zero-recoil limit as the optimal kinematic point at which to measure
$b\to c\ell\nu$ decays now dominates the extractions of \avcb\ in $B$
physics -- not because it is experimentally optimal (quite the
opposite!)  but because it offers a way to minimize the complications of
strong interactions. If we had similar strategies that would allow us to
extract \avub\ from $b\to u\ell\nu$ measurements without form factor
uncertainties, \avtd\ from $B_d$ mixing measurements without decay
constant and bag parameter uncertainties, or \avts\ from $B_s$ mixing
measurements (or limits) without its corresponding decay constant and
bag parameter uncertainties, we would be well on our way to
understanding the CKM matrix at the few percent level.

In the current state of the field this is an unrealized dream.
Across the spectrum of heavy flavor physics the study of weak-scale
phenomena and the extraction of quark mixing matrix (CKM) parameters
remain fundamentally limited by our
restricted capacity to deal with the strong interaction dynamics.

Moreover, as we look to the future beyond the Standard Model, and beyond
the realm of today's heavy flavor physics, we anticipate that the larger
theory in which the Standard Model lives will certainly be either
strongly coupled or will have strongly coupled sectors.  Both
Technicolor, which is modeled on QCD and is {\it ab initio} strongly
coupled, and Supersymmetry, which needs strongly coupled sectors to
break the supersymmetry, are prime examples of candidates for physics
beyond the Standard Model.  Strong coupling is a phenomenon to be
$expected$: weak coupling is the exception in field theory, not the norm.
Nevertheless our ability to compute reliably in a strongly coupled
theory is far from developed, as evidenced by the careful identification
and exploitation of golden modes in heavy quark physics.  Techniques such
as lattice gauge theory that deal squarely with strongly coupled
theories will eventually determine our progress on all fronts of
particle physics.  At the present time absence of adequate theoretical
tools significantly limits the physics we can obtain from heavy
quark experiments.

Recent advances in Lattice QCD (LQCD), however, may offer hope.
Algorithmic advances, and, to a lesser extent, improved computing
hardware have produced a wide variety of nonperturbative results with
accuracies of order 10-20\%. This is particularly true for analyses of
systems involving heavy quarks, such as $B$ and $D$ mesons or the
$\Upsilon$ and $\psi$ quarkonia. First-generation unquenched
calculations have been completed for decay constants and semileptonic
form factors, for mixing and for spectra.  There is strong interest
within the LQCD community in pursuing much higher precision, and the
techniques needed to reduce errors to a few percent exist. 
A small number
of calculations have achieved errors of 5\% or less, including
calculations of heavy-quark masses and the strong coupling constant;
much more is possible within the next few years. But the push towards
high precision is hampered by a lack of sufficiently accurate data
against which to test and calibrate the new theoretical techniques.

\Cleoc\ proposes to address this challenge by confronting it in the charm system
at threshold where the experimental conditions are optimal.  With high
statistics data obtained from the decays of charmed mesons and
charmonium, we will provide unprecedentedly precise data to confront
theory. We will supplement the charmonium data with $\sim 4$ \fbinvv\ of
bottomonium data to be taken by CLEO III in the year prior to conversion
to \Cleoc.  Decay constants, form factors, spectroscopy of open and
hidden charm and hidden bottom, and an immense variety of absolute branching
ratio determinations will be provided with accuracies at the level of
1-2\%.  Precision measurements will demand precision theory.

The measurements proposed below are therefore an essential and
integral part of the global program in heavy flavor physics of this
decade and the larger program of the as yet unknown physics of the next
decade.  By exploiting capabilities which are unique to the charm sector
and the charm energy region, and programmatic opportunities that are
unique to CESR and CLEO, our measurements will explore a large set of
critical weak and strong interaction phenomena.  These in turn will
drive theoretical advances that will both extend and enable the full
program of flavor physics targeted by Babar, Belle, CDF, D0, BTeV, LHCb,
ATLAS, and CMS, and will lay the foundation for strong interaction theory to
meet the requirements of future physics beyond the Standard Model.

\subsection{ Run Plan and Datasets}

The CESR accelerator will be operated at center-of-mass energies
corresponding to  $\sqrt{s} \sim 4140$, $\sqrt{s} \sim 3770 
(\psi^{\prime\prime})$, and $\sqrt{s} \sim 3100 (\Jpsi)$ for
approximately one calendar year each.  Taking into account the
anticipated luminosity which will range from $5\lumi{32}$ down
to about $1\lumi{32}$ over this energy range, the proposed
run plan will yield $3~\fbinvv$ each at the $\psi^{\prime\prime}$ and at
$\sqrt{s} \sim 4140$ above $D_s\overline{D_s}$ threshold,
and $1~\fbinvv$ at the $\Jpsi$.  These integrated luminosities
correspond to samples of 1.5 million $\ds\dsbar$ pairs, 30 million $\dd\dbar$ pairs,
and one billion $\psi$ decays. As a point of reference, note that
these datasets will exceed those of the Mark~III experiment by factors
of 480, 310, and 170, respectively.  If time and luminosity allow,
modest additional data samples will be obtained at the
$\Lambda_c\overline{\Lambda_c}$ threshold region, the
$\tau^+\tau^-$ threshold region, the $\psi(3684)$, and over
a set of scan points for an $R$ vs. $\sqrt{s}$ determination

In addition, prior to the conversion to low energy operation, we plan to
take $\sim 4$ \fbinvv\ spread over the $\Upsilon(1S)$, $\Upsilon(2S)$,
and $\Upsilon(3S)$ resonances to launch the QCD part of the
program. These datasets will increase the available $b\bar b$ bound
state data by more than an order of magnitude.

\subsection{Hardware Requirements}

The conversion of the CESR accelerator for low energy operation
will require the addition of 18 meters of wiggler magnets to
enhance transverse cooling of the beam at low energies. In the CLEO III
detector the solenoidal field will be reduced to 1.0 T,  and 
the silicon vertex detector may be replaced with a small, low mass
inner drift chamber.

No other changes are needed to carry out the proposed program.

\subsection{Measurements}

The principal measurement targets include:

\Begitem
\item
{\bf Leptonic charm decays:} $D^-\to \ell^-\overline{\nu}$ 
and $D_s^-\to\ell^-\overline{\nu}$.\\  From
the muonic decays alone the decay constants $\fd$ and
$\fds$ can be determined to a precision of about 2\%.  The decay constants measure the
nonperturbative wave function of the meson at zero inter-quark
separation and appear in all processes where constituent quarks must
approach each other at distances small compared to the meson size. 
Note that while $f_\pi$ and $f_K$ are known to 0.3\% and 0.9\% respectively,
\fds\ and \fd\ are only known to about 35\% and
100\% respectively, and \fb\ and \fbs\ are unlikely to be measured to
any useful precision in this decade.

\item
{\bf Semileptonic charm decays:} $D\to (K,K^*)\ell\nu$, $D\to
(\pi,\rho,\omega)\ell\nu$, $D_s\to (\eta, \phi)\ell\nu$, $D_s\to
(K,K^*)\ell\nu$, and $\Lambda_c\to\Lambda\ell\nu$.\\  Absolute branching
ratios in critically interesting modes like $D\to\pi\ell\nu$ and $D\to
K\ell\nu$ will be measured to $\sim 1\%$, and form factor slopes to
$\sim 4\%$. Form factors in all modes can be measured across the
full range of \qsq\ with excellent resolution. Semileptonic decays are
the primary source of data for the CKM elements \avub, \avcb,
\avcd, and \avcs, but these CKM elements cannot be extracted without 
accurate knowledge of form factors.  Currently, semileptonic branching
ratios are known with uncertainties that range from 5\% to 73\% -- in
the cases where they are known at all -- and form factor measurements
are limited by resolution and background. Inclusive semileptonic decays
such as $D\to eX$, $D_s\to eX$, and $\Lambda_c\to eX$ will also be
examined and branching ratios will be measured to a precisions
of $1-5\%$.  Currently, such quantities are known with uncertainties that
range from 4\% to 63\%.

\item
{\bf Hadronic decays of charmed mesons.}\\ 
The rate for the critical normalizing modes $D\to K\pi$, $D^+\to
K\pi\pi$, and $D_s\to\phi\pi$ will be established to a precision of order $1-2 \%$. Currently
these are known with uncertainties that range up to 25\% and are even
larger for other hadronic decays of interest. Many important $B$ meson branching
ratios are normalized with respect to these subsidiary charm modes.

\item
{\bf Rare decays, $D\dbar$ mixing, and CP violating decays.}\\ 
\Cleoc\  can search for rare decays with a typical sensitivity of
$10^{-6}$, study mixing with a sensitivity to $x=\Delta M /M$ and
$y=\Delta \Gamma/2\Gamma$ of under 1\%, and detect any CP violating
asymmetries that may be present with a sensitivity of better than 1\%.
\Cleoc\ will also search for evidence of new physics within $\tau$ decays.

\item
{\bf Quarkonia and QCD.}\\ With approximately
one billion $\Jpsi$'s produced, \Cleoc\ will exploit the natural glue factory,
$\psi\to\gamma gg\to\gamma X$, to search for ``glueballs'' and other glue-rich
states. The region of $1 < M_X < 3$ GeV/$c^2$ will be explored with
partial wave analyses for evidence of scalar or tensor glueballs,
glueball-$q\overline q$ mixtures, exotic quantum numbers, quark-glue
hybrids, and other evidence for new forms of matter predicted by QCD but
not yet cleanly observed.
\Begitem
\item
Masses, widths, spin-parity quantum numbers, decay modes, and production
mechanisms will be established for any states that are
identified.
\item
Reported glueball candidates such as the tensor candidate
$f_J(2220)$, and the scalar states $f_0(1710)$, $f_0(1500)$,
and $f_0(1370)$ will be explored in detail and spin-parity assignments
clarified.
\item
The inclusive photon spectrum in $J/\psi\to\gamma X$ will be examined with 
$<20$
MeV photon energy resolution.  States with up to 100 MeV width and
inclusive branching ratios above $1\times 10^{-4}$ will be
identified.
\Enditem
The $\sim 4$ \fbinvv\ of CLEO $b\bar b$ resonance data
(to be taken prior to conversion to low energy operation) 
will also be exploited to survey the
physics of the $\Upsilon(1S)$, $\Upsilon(2S)$, and $\Upsilon(3S)$,
resonances. We will measure leptonic widths (related to decay
constants of mesons with open flavor) and photonic transition matrix
elements (related to form factors in semileptonic decays of open
flavor mesons).  Comparing experimental results with LQCD predictions
for the $\Upsilon$ (and $\psi$) spectra, leptonic widths and form
factors test both the heavy-quark action that is used for LQCD
simulations of $B$'s and $D$'s, and the specific techniques used to
analyze $B$ and $D$ decay constants and form factors in LQCD.  \Cleoc\ will
also make spectroscopic searches for new states of the $b \bar b$ system
and for exotic hybrid states such as $cg\bar c$ and perhaps $bg\bar b$.
Analysis of $\Upsilon(1S)\to\gamma X$ will play an important role in
establishing or debunking any glueball candidates found in the $J/\psi$
data.

\item
{\bf Spot checks of $R$.}\\ The ratio $R=\sigma(e^+e^-\to{\rm hadrons})/
\sigma(e^+e^-\to \mu^+\mu^-)$ will be measured at various values of $\sqrt{s}$
with a precision of 2\% per point. The
$R$ measurements are critical to interpretation of precision
electroweak data and the $g-2$ experiment.

\Enditem

\subsection{Unique Features of the \Cleoc\ Program}

Many of the measurements described above have been done or attempted by
other experiments such as Mark~III and BES, and many are accessible to
the B-factory experiments operating at the $\Upsilon(4S)$.  What makes
\Cleoc\ unique?

Compared to the Mark~III and BES experiments which have taken data on the
same $\psi$ resonances as we propose here, \Cleoc\ will have:
\Begitem

\item
{\bf Vastly more data.}\\  As noted above the \Cleoc\ data sample will be $\sim 200-500$
times larger than the corresponding Mark~III datasets.  Compared to BES,
\Cleoc\ will have 270 times as much $D$ and $D_s$ data, and 20 times as many
$\psi(3100)$ decays.  One order of magnitude opens new vistas; two
orders of magnitude can change a field.

\item
{\bf A modern detector.}\\ Mark~III was built twenty years ago, and BES
was modeled on Mark~III.  Detectors have gone through several
generations of development since then. In every resolution and
performance parameter -- hit resolution, momentum and energy resolution,
mass resolution, particle ID capability, solid angle coverage -- CLEO
III is superior to these other detectors by substantial margins. Photon
energy resolution, for example, is factors of 10-20 times better
(depending on $E_\gamma$); charged particle momentum resolution is 2-3
times better (depending on $p_T$).  Particle identification with the
RICH detector, augmented by 
energy loss ($dE/dx$) measurements in the drift 
chamber,  give tens to hundreds of sigma $K\pi$ separation across
the full kinematic range. A $25\%$ increase of solid angle coverage
relative to BES gives \Cleoc\ a huge advantage in analyses such
as double-tag measurements that require every particle to be
reconstructed.  The gains go as $1.25^n$ where $n$ is the total
track and photon multiplicity of the event. This implies a typical
effective luminosity gain of 8 in such analyses. For studies that
involve partial wave analysis, the increase in solid angle coverage
means angular distributions can be measured across the full angular
range without large variations in efficiency.  This translates to
substantial gains in the reliability and precision with which $J^{PC}$
can be measured for a given state.

\Enditem

On the other hand, \Cleoc\ will $not$ have any advantage in statistics or
in detector performance when compared to Babar and Belle.  The three
detectors are all similar, and with an anticipated 400 \fbinvv\ of
$\Upsilon(4S)$ data, Babar and Belle will each have about 500 million 
continuum $e^+e^-\to c\bar c$ events. Yet the data \Cleoc\ takes at charm
threshold has distinct and powerful advantages
over continuum charm production data taken
at the B-factories, which we list here:

\Begitem
\item
{\bf Charm events produced at threshold are extremely clean.}\\  The
charged and neutral multiplicities in $\psi(3770)$ events are 5.0 and
2.4, compared with 11.0 and 5.6 in $\Upsilon(4S)$ events.  This alone
substantially reduces combinatorics, but in addition the $\psi(3770)$
decays are spherical, distributing decay products uniformly in the
detector solid angle.  Low multiplicity in \Cleoc\ translates to high
efficiency and low systematic error.

\item
{\bf Charm events produced at threshold are pure $D\bar D$.}\\  No
additional fragmentation particles are produced.  The same is true for
$\psi(4140)$ decaying to $D \bar{D^*}$, $D_s \bar{D_s}$ and $D_s
\bar{D_s}^*$, and also for threshold production of $\Lambda_c
\bar{\Lambda_c}$.  This allows the use of kinematic constraints such as
total candidate energy and beam constrained mass, and also permits
effective use of missing mass methods and neutrino reconstruction. The
crisp definition of the initial state is a uniquely powerful advantage
of threshold charm production that is absent in continuum charm
production.

\item
{\bf Double-tag studies are pristine.}\\ The pure production of
$D\bar D$ states, together with the low multiplicity and high branching
ratios characteristic of typical $D$ decays permits effective use of
double-tag studies in which one $D$ meson is fully reconstructed and the
rest of the event is examined without bias but with substantial
kinematic knowledge. These techniques, pioneered by Mark~III many years
ago allow one to make $absolute$ branching ratio determinations.
Backgrounds under these conditions are heavily suppressed.
Very low background conditions
minimize both statistical and systematic errors.

\item
{\bf Signal/Background is optimum at threshold.}\\ The
cross section for the signal $\psi(3770)\to D\bar D$ is equal to
the cross section for the underlying continuum $e^+e^-\to {\rm hadrons}$
background.  By contrast, for $c\bar c$ production at $\sqrt{s}=10.6$ GeV the
signal is only 1/4 of the total hadronic cross section.  In addition,
the $c\bar c$ fragmentation distributes the final states
among many charm hadron species.

\item
{\bf Neutrino reconstruction is clean.}\\ For leptonic and semileptonic
decays the lost neutrino can be treated as a missing mass problem and in
the double tagged mode these measurements have low backgrounds.  The
missing mass resolution is under a pion mass.  For
semileptonic decays this also means that the resolution on \qsq\ is
excellent, about 3 times better than is available in continuum charm
reconstruction at $\sqrt{s}=10.6$ GeV.  


\item
{\bf Quantum coherence.}\\ For $D$ mixing and some CP violation studies,
the fact that the $D$ and $\bar D$ are produced in a coherent quantum
state in $\psi(3770)$ decay is of central importance for the subsequent
evolution and decay of these particles. The same is true for the $CP=+1$
mode $\psi(4140)\to\gamma D \bar D$.  The coherence of the two
initial-state particles allows simple methods to measure $D \bar
D$ mixing parameters and check for direct CP violation.

\Enditem 

In addition to the advantages of studying open-charm decays at
threshold, the \Cleoc\ program includes the opportunity to use a huge
charmonium data sample in searches for glueballs, hybrids, and exotic
states.  If found -- or if $not$ found -- these states will present a
powerful challenge to QCD calculations. Furthermore, CLEO will have the
unique ability to compare results between both high statistics $J/\psi$
$and$ $\Upsilon$ data sets, and further cross check with the 25 \fbinvv\
of existing two-photon data. These corroboratory measurements will be
used to eliminate spurious glueball candidates. Theory, moreover, will
be forced to confront precision data in both open- and hidden-flavor
charm and bottom mesons simultaneously.

Taken together, these technical and programmatic features constitute
formidable advantages for the \Cleoc\ proposal. 

\section{The Impact of \Cleoc}

The measurements of leptonic decay constants and semileptonic form
factors, together with the study of QCD spectroscopy both in the
$c\bar c$ and $b\bar b$ quarkonium sectors will yield an extensive set of
$1-2\%$ precision results that will rigorously constrain theoretical
calculations.  The calculations which survive these tests will be
validated for use in a wide variety of areas where the interesting
physics cannot be extracted without theoretical input.  This
broader impact of \Cleoc\ results extends beyond the borders
of \Cleoc\ measurements and affects most of the core issues in
heavy flavor physics.  We list here some of the areas that will
be most notable:

\Begitem
\item
{\bf Extraction of \avub.}\\ Currently limited by form factor calculations
to an estimated 25\% accuracy, and unlikely to improve beyond 10\% in
present-day extrapolations.  Pinning down form factor technology in 
the closely related charm decays such as $D\to\pi\ell\nu$ and
with $D\to\rho\ell\nu$, \Cleoc\
data will open the door to 5\% or better precision in \avub. 

\item
{\bf Extraction of \avtd\ and \avts.}\\  Currently limited by ignorance of
$\fb\sqrt{B_{B_d}}$ and ${\fbs} \sqrt{B_{B_s}}$, our only prospect for
separately extracting \avtd\ and \avts\ from $B$ mixing
measurements is through improving decay constant calculations to the
percent level.  Determinations of the charmed decay constants
\fd\ and \fds\ will underwrite the required theoretical advances and
open the door to $\sim5\%$ determinations of \avtd\ and \avts.

\item
{\bf Extraction of \avcd\ and \avcs.}\\  Currently known only at the 
$\sim 10\%$ level by direct measurement, \Cleoc\ will provide absolute
branching ratio measurements of leptonic and semileptonic decays from
which \avcd\ and \avcs\ can be extracted to $\sim$1\% accuracy.  Here as
elsewhere, form factor and decay constant calculations must be advanced
to a comparable level of precision and validated by the entire range of \Cleoc\
measurements for CKM determinations at the percent level to be valid.

\item
{\bf Extraction of \avcb.}\\
Currently limited by a variety of both experimental and theoretical
inputs. Prominent among these are the theoretical
control of the form factors and the experimental determination of
${\cal B}(D\to K\pi)$. \Cleoc\ will drive form factor technology,
and measure the normalizing branching ratios at the sub-percent level.

\item
{\bf Unitarity tests of CKM.}\\  Currently poorly satisfied by the first
two rows of the CKM matrix, which fail both orthogonality and
orthonormality conditions at the $2-3\sigma$ level.  Unitarity
conditions can be probed with 1\% precision when \avcd\ and \avcs\ are
provided at this level by \Cleoc.

\item
{\bf Over-constraining the Standard Unitarity Triangle.}\\  Provided \avub\
and \avtd\ have been determined at the 5\% level, as discussed in
items 1 and 2 above, the triangle sides will have been
measured with precision comparable to the phase quantity $\sin2\beta$ --
thereby allowing for the first time meaningful comparison of the sides
of the unitarity triangle with one of the angles.  

\item
{\bf New forms of matter.}\\
In the quarkonium studies new forms of gluonic matter may be identified.
Current results in this field are murky and contradictory. The
high statistics data sample, high resolution detector, and clean initial
state will be an unprecedented combination in this field, and offer the
best hope for incisive experimental results.

\Enditem

\section{Measurements of $R$} 

\Cleoc\ can approach measurement of $R$ in two ways.  The first
involves an explicit scan over $\sqrt{s}$ to measure $R$ directly
at different energies.  The second involves the use of radiative returns
to determine an average $R$ over a range of energies below those
accessible directly by CESR-c.

\subsection{$R$ scanning}

A scan of 
the energy range between 3 and 5 GeV will clarify the energy dependence of $R$
just below the open charm threshold, from $J/\psi$ to $\psi(3770)$,
and above it where several relatively broad $c\bar{c}$
resonances exist. A possible scenario is to scan this energy
range with a steps of 100 and 20 MeV, below and above
$\psi(3770)$ respectively, collecting about $10^4$ hadronic events per
point. 
Such a scan will require an integrated luminosity of about 
$100~\rm{pb}^{-1}$.
It will also serve as an introduction to a more detailed potential study 
of the $c\bar{c}$ resonances in this energy range as well as 
$D^*$ and $D_s$ mesons copiously produced in their decays 

The previous experience of CLEO 
(which measured $R$ with an accuracy of 2\% 
in the vicinity of the $\Upsilon(4S)$ resonance \cite{Ammar98}) and 
recent progress in the calculations of radiative corrections together 
with the hermeticity of the CLEO-III detector allows one to expect a 
systematic uncertainty of about 2\%. After that a scan of the energy 
range from 5 to 7 GeV will be needed to solve the dramatic contradiction 
between the old measurement of MARK-I \cite{MARK-I} and the unpublished 
results of Crystal Ball \cite{CBALL}. Here a scan in steps of 100 MeV with 
$10^4$  events per point will be adequate,
requiring an integrated luminosity of $50~\rm{pb}^{-1}$. At
an average luminosity of $3\times10^{32}~\rm{cm}^{-2}\rm{s}^{-1}$
it will take one week of collection time to scan the entire energy range between 3 and 7
GeV.

\subsection{Radiative returns}

The second approach to measuring $R$ at \Cleoc\ will be to use radiative
return events \cite{BEN99,KUHN}. Such events would allow \Cleoc\ to measure 
$R$ in the 1-3 GeV energy range which is of crucial importance to reduce the
overall uncertainty in $\alpha(M_Z)$ and $(g-2)_{\mu}$.
Recent experience of BABAR confirms statistical feasibility of such
$R$ measurements \cite{solodov}. While running at the $\psi(3770)$ 
one can expect $\sim 10^{4}$ fully contained radiative return events 
in the range 1-3 GeV 
per $1~\rm{fb}^{-1}$ of data. 
However, it is still unclear whether systematic errors can be controlled 
well enough to make the radiative return measurements meaningful, \ie, 
to reach an uncertainty of less than 5\%.

Various dedicated experiments have recently been proposed to
improve our knowledge of $R$ in the crucial 1--3 GeV energy range
\cite{VENAN}. If they are approved, the \Cleoc\ measurements
would be most valuable as an independent check of the dedicated
experiments.

A 2\% measurement of $R$ between 3 and 7 GeV will significantly 
reduce the uncertainty of the hadronic contribution to
$(g-2)_{\mu}$ and especially $\alpha(M_Z)$. 
If radiative return events can be utilized to reach the 5\% accuracy 
between 1 and 3 GeV, remarkable improvement is expected. For example,
the $\alpha(M_Z)$ uncertainty would be a factor of 2 smaller than 
it is at present. 

\subsection{`Modern' R Measurements for Charm Spectroscopy Studies}

Although $c\bar{c}$ states below strong decay threshold
(3.7 GeV) were studied at $e^+e^-$ machines some time ago, higher
mass states are very poorly defined.  Most quark model 
analyses~\cite{ClosePage} 
find 3 poorly defined L=2 states above a mass of 4 GeV based on
on low statistics $R$ data from 30 years ago.  A detailed study of 
the properties of these
states will test the long-range $q\bar{q}$ potential in a controlled
way.  In addition, the mass
region from 4-5 GeV is where lattice calculations have predicted
charmed hybrids to be, as detailed in other sections of this document.

Thus, a means to study this spectroscopy with more detail
than the previous experiments is required.  \Cleoc\ proposes
a study that will provide a direct look at the states through
a scan of the mass region 3.7-5 GeV.  Because very little is known 
about the states above $D\bar{D}$ threshold, a broad look is 
appropriate.  However, the complexities of extraction above the inelastic
threshold require more detailed information about the final state.
The proposed study will be similar to the $R$ measurement described
above.  However it will measure the final state whenever possible
by reconstructing $D$ mesons in that the dominant
decay modes will be $D\bar{D}$, $D^*\bar{D}$, $D^*\bar{D^*}$, 
and $D^{**}\bar{D}$.  Previous measurements of these decay
rates are strongly in disagreement with quark model 
calculations~\cite{ClosePage}, but the data is of poor quality.

Data would be collected with a loose trigger in the region 3.7-5 GeV.
Similar to an $R$ measurement, runs will be at closely spaced
energies (steps of 20 MeV).  However, the runs need to be much longer
($\sim$100,000 events) than is typical~\cite{bes_r}.
Although the BES data have higher statistics
than the older SLAC experiments, no attempts have been made
to analyze the composition of the final state because of 
detector limitations.  With the larger event samples and the 
acceptance of CLEO, one would be able to reconstruct at least
one $D$ meson in about 10\% of the events.  This would be
sufficient to characterize the final state and measure
the angular distribution.  The overall efficiency (summed decay 
branching ratio times detection efficiency) is about 16\% for 
$D^0$ mesons and 6\% for $D^+$ mesons for prominent decay 
modes.  
At a luminosity of $3 \times 10^{32}$ cm$^{-2}$s$^{-1}$, each step
would require about 2 days of data taking.

Existing quark model calculations for charmed hybrid 
mesons~\cite{BarnesClosePageSwanson} assume the quarks to have 
orbital angular momentum 1.  For decay mechanisms normally 
used in quark models, decays to $D\bar{D}$ mesons will be suppressed.
If this symmetry is correct, a state with very low decay
rate to $D\bar{D}$ would signal the possibility of a
hybrid.  Total widths are about 10-30 MeV in these models, 
similar to the conventional $c\bar{c}$ states at the same
mass.

This measurement is related to other \Cleoc\ measurements outlined
in this document.  It is closely related to the $R$ measurement
discussed in the previous section, but will require much larger
statistics.

\section{Summary}

The high-precision charm and quarkonium data we propose to
take will permit a broad suite of studies of weak and strong
interaction physics.  In the threshold charm sector measurements are
uniquely clean and make possible the unambiguous determinations of
physical quantities discussed briefly above, and at greater length in
the chapters that follow. The advances in strong interaction
calculations that we expect to drive will in turn underwrite advances in
weak interaction physics not only in \Cleoc, but in all heavy quark
endeavors and in future explorations of physics beyond the Standard
Model.

\Cleoc\ stands to make a significant impact upon the $R$ determination
in  3-7 GeV range, with direct 2\% measurements appearing feasible
with a fairly detailed scan.  In addition, the use of radiative returns may
provide a useful crosscheck for direct low energy measurements at
other facilities if the systematic uncertainties can be controlled at
the 5\% level.

\end{document}